# Chapter 10: Aortic and arterial mechanics

Author(s):

*S. Avril (Université de Lyon, IMT Mines Saint-Étienne, France, avril@emse.fr)

## 10.1. Introduction.

Knowledge of the mechanical properties of the aorta is essential as important concerns regarding the treatment of aortic pathologies, such as atherosclerosis, aneurysms and dissections, are fundamentally mechanobiological. A comprehensive review is presented in the current chapter. As the function of arteries is to carry blood to the peripheral organs, the main biomechanical properties of interest are elasticity and rupture properties. Regarding elastic properties, an essential parameter remains the physiological linearized elastic modulus in the circumferential direction, although sophisticated constitutive equations including hyperelasticity are available to model the complex coupled roles played by collagen, elastin and smooth muscle cells in the mechanics of the aorta. Regarding rupture properties, tensile strengths in the axial and circumferential directions are on the order of 1.5 MPa in the thoracic aorta and the radial strength, which is important regarding dissections, is about 0.1 MPa. All these properties may vary spatially and change with the adaptation of the aortic wall to different conditions through growth and remodelling. The progression of diseases, such as aneurysms and atherosclerosis, also manifests with alterations of these material properties. After presenting how the mechanical properties of elastic arteries, including the aorta, can be measured and how they can be used in models to predict the biomechanical response of same under different circumstances, a section will focus on the biomechanics of the ascending thoracic aorta and the chapter will end with current challenges regarding predictive numerical simulations for personalized medicine.

## 10.2. Mechanical properties of arteries

### 10.2.1. Mechanical function of arteries

#### 10.2.1.1. Arterial pressure and wall stress

Arteries are in charge of carrying blood to the peripheral organs. The blood pressure $P$ exerts a force perpendicular to the luminal surface of the artery, which causes its distension. It is compensated by radial and circumferential wall tensions that oppose distension. Assuming a perfectly cylindrical shape, the resulting circumferential stress $\sigma_{circ}$ can be estimated by Laplace's law, considering the wall as a thin, long circular tube of radius $r$ and thickness $t$. There are several expressions of Laplace's law, such as $\sigma_{circ} = Pr/h$ [1], according to which the circumferential stress is directly related to the pressure and to the geometry of the wall. It is also accepted that modifications of the circumferential stress lead to a remodeling of the wall and changes in its structure. In particular, these changes include thickening of the wall (hypertrophy of the smooth muscle) when mechanical stress is increased; conversely, wall atrophy when it is decreased [2].

Let us note, however, that this estimate by Laplace's law gives only an average value. It neglects the influence of the microstructure of the wall and does not make it possible to evaluate the transmural variations of the circumferential stress. In addition, blood pressure is not constant, but varies during the cardiac cycle. In particular, the pulse pressure plays an important role in the remodeling of large arteries [3].

Figure 1 shows two successive cycles of arterial pressure. There is a rise in pressure during systolic ejection (1), followed by a peak (2) and then a decrease in pressure during ventricular diastole (3). The measured systolic blood pressure (SBP) is the maximum value at the peak. The dicrotic notch (4) corresponds to the closure of the aortic valve and is followed by a drop in diastolic pressure (5). The measured diastolic blood pressure (DBP) is the minimum value of the curve (6). These characteristic values of blood pressure can be measured with an automatic or manual sphygmomanometer with stethoscope. The pressure can also be recorded continuously over several cycles, either by means of a catheter inserted in the artery, or by a non-invasive technique, such as applanation tonometry, when the access is possible (common carotid arteries for instance). Finally, it should be emphasized that the arterial pressure curve is not the same according to the location of the artery analyzed in the vascular tree. It depends on the size and elasticity of the artery in question. Normal blood pressure values in humans are between 100 and 140 mmHg for SBP and between 60 and 90 mmHg for DBP. The mean arterial pressure (MAP) corresponds to the area under the pressure curve. It is closer to DBP than to SBP and is assumed to be nearly constant throughout the arterial tree in the supine position, at least until blood reaches the resistance arterioles.

### 10.2.1.2. Arterial compliance

The heart acts as a pulsatile pump that propels blood into the vascular system during its contraction (systole). Part of the blood ejected during systole is stored during distension of the aorta and the proximal arteries. The function of the large elastic arteries is to relay the contraction of the heart when it enters its relaxation phase (diastole), thanks to their compliance. After closing the aortic valve, the aorta and the proximal arteries retract elastically and restore the volume of blood stored. This is the Windkessel effect through which blood pressure is maintained and blood flow is increased in diastole, which ensures a continuous and non-pulsatile flow in the peripheral arteries and the capillaries. This function of diastolic relay of the cardiac contraction is directly related to the elastic properties of the large arteries, which are conferred on them by the large quantity of elastic fibers and type III collagen in the media [1,2].

The compliance of an artery can be measured experimentally from the pressure-volume relationships. Indeed, arterial compliance is defined as the ratio between the increase in luminal volume and the increase in pressure that induces it: it corresponds to the slope of the pressure-volume curve. In addition, arterial distensibility is defined as the relative variation of volume with respect to the pressure variation. It is therefore a relationship between compliance and luminal volume.

Since the pressure-volume relationship is not linear for an artery, compliance and distensibility vary with the pressure level. The higher the pressure, the more the collagen fibers are stressed and the greater the resistance to distension. Thus, compliance and distensibility decrease as pressure increases. It should be noted that these two parameters are often measured by studying a cross-section of the artery, with the assumption that the artery does not change in length. Cross-sectional studies can be carried out using external ultrasound probes for superficial arteries such as common carotid and femoral arteries, and the radial or brachial arteries, or by magnetic resonance imaging (MRI) for the aorta [4].

### 10.2.1.3. Pulse wave

The impact produced by the blood on the walls of the aorta during cardiac ejection generates a pressure wave called a pulse wave which propagates along the arterial tree from the aorta to the peripheral arteries. It causes a radial deformation of the arterial wall, which enables diastolic relay of the cardiac contraction. The pulse wave velocity (PWV) is considered as a parameter representing arterial stiffness. Indeed, the Moens-Korteweg relation relates the PWV to the geometrical and mechanical characteristics of the artery when it is considered as a thin-walled isotropic tube of infinite length:

$$VOP = \sqrt{\frac{E_{inc}h}{2\rho r}} \qquad (10.1)$$

where $E_{inc}$ is the incremental Young's modulus of the arterial wall, $h$ its thickness, $r$ its radius and $\rho$ the density of the blood. The stiffer the artery (i.e. the higher its modulus of elasticity), the higher the PWV. For a healthy middle-aged individual, PWV generally ranges between 4 and 10 m/s.

When the incident pressure wave encounters bifurcations or peripheral resistance arteries, it is partially reflected. The reflected wave propagates backwards and is added to the incident wave. The reflected wave has an important impact on the shape and amplitude of the pressure profiles and explains their variations along the arterial tree. Indeed, in the peripheral arteries located closer to reflection sites, the reflected wave returns more quickly than in other arteries. The reflected wave adds to the incident wave earlier in the cardiac cycle by increasing SBP while it mainly affects DBP for elastic arteries near the heart (Figure 2). In the same way, in the elderly or patients suffering from vascular pathologies and in which the arteries are more rigid, the SBP is increased by the reflected wave. This phenomenon can be quantified by the increase index defined as the ratio between the pressure increase and the pulse pressure, the pressure increase being the difference between the first and second systolic peaks [4].

### 10.2.2. Nonlinear elasticity

The mechanical properties of an artery depend on its geometry and on the proportions of its various microconstituents, described in a previous chapter of this book. Smooth muscle cells (SMCs) control the vasomotion of the arteries and what are called their active mechanical properties while the passive mechanical properties of the arteries are determined by the extracellular matrix (ECM), in particular the elastin and collagen fibers. The elastic arteries of large caliber in which we are interested in this chapter contain more elastic fibers than muscular fibers. They therefore exhibit a small vasomotricity and their passive properties are preponderant in the characterization of their mechanical behavior. The active properties related to the muscular tone of the arterial wall will be discussed in Section 1.7.

Elastin fibers are highly elastic and resist to distension forces generated by blood pressure. Collagen fibers are much more rigid and almost inextensible, and prevent the distension of the artery. However, they are arranged in the wall so as to be stressed only under high distension levels. They are recruited gradually when the pressure increases and are only truly stretched for high pressures, above about 125 mmHg in healthy situations. Thus, the arteries exhibit a highly non-linear pressure-diameter response with characteristic stiffness under high pressure. Individual contributions of elastin and collagen fibers to the overall non-linear response of arterial tissue were identified by Roach and Burton [5]. The difference in behavior between elastin and collagen was demonstrated by recording the tension-circumference relationships of intact human external iliac arteries and then performing selective enzymatic digestions of collagen and elastin.

It appears that the distension of an artery mainly depends at low pressure on its elastic fibers and high pressure on its collagen fibers. The initial slope of the stress-strain curves roughly indicates the state of the elastic fibers while its final slope roughly reflects the state of the collagen.

The arteries therefore have a nonlinear elastic behavior that can be modeled by a hyperelastic behavioral relationship, originally developed for elastomers such as rubber. The stress field is obtained by deriving a deformation energy function of which several types have been proposed in the literature: exponential, polynomial, etc... We will return to this in detail in Section 3.3.

**10.2.3. Incompressibility**

Like many soft biological tissues, the arteries are generally considered to be incompressible. Patel and Vaishnav [6] demonstrated the incompressibility of the arteries experimentally by comparing the resistance of the material to changes in volume. Carew et al. [7] immersed arterial segments in water in a closed chamber and observed a volume change of 0.165% at a pressure of 181 mmHg. They concluded that this result justified the assumption of incompressibility. Finally, Tardy [8] measured in-vivo the relative variations in volume per unit length of the radial artery and found values below 1% under physiological conditions.

**10.2.4. Anisotropy**

Numerous studies agree that the vascular tissue is anisotropic, that is to say that its elastic properties are not the same in all directions. The first study documenting the anisotropy of the arteries was carried out by Patel and Fry in 1966 on dog arterial segments [9]. Several studies have subsequently confirmed this result [9-13].

Nevertheless, many authors have modeled arteries using isotropic constitutive equations. Dobrin and Doyle (1970) showed that, by assuming isotropy, the circumferential elastic modulus was overestimated by 17% at a pressure of 80 mmHg [11].

**10.2.5. Residual stresses**

Intuitively, the reference state chosen to study an arterial segment corresponds to a state without external loads, that is to say without pressure or longitudinal forces. But this state does not correspond to a state free of constraints. Indeed, residual longitudinal and circumferential stresses remain in the arterial segment. They are non-uniform and result from tissue growth and adaptation to their mechanical environment during development [14]. They are released only when the vessel is severed and removed from its environment.

The existence of residual circumferential and radial stresses has been demonstrated by Vaishnav and Vossoughi [15] and Fung [16]. They observed that by radially cutting an arterial ring, the artery opens by itself by an angle which can be measured by joining the middle of the internal arc at its extremities (Figure 3). The state "zero stress" can therefore be represented by an open arterial segment. It should be noted that the opening angle varies according to species, the arterial site considered and its condition. Fung and Liu [17] observed changes in opening angle after only a few days of pressure increase (HBP). The opening angle thus appears as a marker of the remodeling of the arterial wall.

It is important to note that residual stresses play a central role in vascular mechanics because they modify the distribution of stresses in the arterial wall [18]. Indeed, they make it possible to reduce the circumferential stresses toward the intimal portion of the wall, as well as the stress gradient across the wall [19]. This result was illustrated by Delfino et al. [20] using an isotropic 3D finite element model of the human carotid bifurcation. Semi-analytical approaches considering the anisotropy of arterial tissue showed the same trend [22, 65]. However, residual stresses as revealed by sometimes extreme opening angles in the elderly may lead to increased stresses towards the adventitial portion of the wall [175].

On the other hand, Greenwald et al. [23] showed that the opening angle is not correlated with smooth muscle activation and the collagen content in the wall, but is rather associated with elastic fibers. Indeed, treatment with elastase on segments of rat aorta reduced the angle of opening while collagenase had no effect. Moreover, the opening angles of the different aortic layers of the wall were different [24].

Azeloglu et. al. [7] investigated, both numerically and experimentally, the regulating role of proteoglycans present in the arterial wall. They showed that, with an inhomogeneous distribution of proteoglycans through the wall thickness, the Donnan osmotic pressure would vary through the wall thickness, resulting in an inhomogeneous swelling stress field in the solid matrix, which would significantly affect the opening angle observed experimentally. The residual stress state in arteries has also been shown to be location-dependent and to be influenced by a host of other factors, a review of which can be found in [5].

### 10.2.6. Axial stretch

During excision of an arterial segment, a reduction in length or longitudinal retraction, is observed. This phenomenon shows the existence of a longitudinal pre-stress exerted by the surrounding tissue. Moreover, Han [25] noted that the longitudinal extension does not vary after the radial opening of the vessels and is therefore the same at zero stress. In-vivo, because of the presence of perivascular tissues, arteries retain a fixed length which does not vary with pressure. Van Loon [26] showed that, in dog arteries, the axial force-length curves obtained in-vitro for different pressure levels intersected at a single point. The length at this point corresponds to a constant axial force regardless of pressure. In turn, this makes it possible to estimate the longitudinal extension in-vivo, defined as the ratio between the length of the arterial segment in-vivo and the length of the same in the reference state without constraint [90].

### 10.2.7. Vascular contraction

A unique feature of many soft tissues is their ability to contract via actin–myosin interactions within specialized cells called myocytes. Examples include the cardiac muscle of the heart, skeletal muscle of the arms and legs, and smooth muscle, which is found in many tissues including the airways, arteries, and uterus. A famous equation in muscle mechanics was postulated in 1938 by A.V. Hill to describe the force–velocity relationship. This relationship, like many subsequent ones, focuses on the 1-D behavior of the myocyte or muscle along its axis; data typically comes from tests on muscle fibers or strips, or in some cases rings taken from arteries or airways. Although much has been learned, much remains to be discovered particularly with respect to the multiaxial behavior. The interested reader is referred to Fung [16].

### 10.2.8. Vascular adaptation

Vascular adaptation refers to the cell-mediated mechanical effects of growth, atrophy, remodeling, and healing. Like all tissues and organs, the vasculature undergoes many changes during normal development and aging as well as due to disease and injury. Morphological, histological, and biomechanical changes stem from growth and remodeling that follow maturation. Under normal conditions, the mature arterial wall is fairly stable: turnover of endothelial cells is on the order of 0.02% per day, except in regions of complex flow (e.g., near a bifurcation) where it may be as high as a few percent per day; turnover of smooth muscle cells is about 0.06% per day, and the half-life of collagen is on the order of weeks to months, whereas the half-life of elastin is comparable to the life span of the organism [133]. Normal tissue maintenance is thus accomplished via slow, steady processes by which constituents are turned over in a balanced manner, the deposition of proteins being balanced by degradation. The natural tendency toward stability of normal biological states is called *homeostasis*. In response to persistent nonhomeostatic stimuli, as altered needs of distal tissues due to consistent exercise or inactivity, local alterations in hemodynamic loads, local changes in the expression of growth factors or vasoactive substances, microgravity, and surgery, the rates of turnover of various constituents can increase dramatically and the arterial wall can quickly undergo significant modification via unbalanced cellular activity: hyperplasia, hypertrophy, apoptosis, and migration of cells as well as via unequal synthesis and degradation of extracellular matrix. The biomechanical theory for modeling these effects will be briefly explained in Section 3.5.

### 10.2.9. Alterations of mechanical properties in aging and vascular pathologies

There are many different types of arterial pathologies and they together represent one of the leading causes of death in the world. Myocardial infarction and stroke affect 120,000 and 150,000 people per year in France, respectively. Stroke is the leading cause of disability in adults worldwide, the second leading cause of dementia after Alzheimer's disease and the third leading cause of death in industrialized countries after heart disease and cancer. Myocardial infarction and stroke are mainly triggered by the obstruction of an artery feeding respectively the heart or the brain. This leads eventually to ischemia, defined as the metabolic demand being unmet, leading to tissue suffering and ultimate necrosis. Ischemia is most often triggered by the rupture of an atheroma plaque, a lipid deposit that forms on the inner wall of the arteries with age and under the influence of various risk factors (sex, heredity, food, physical inactivity, smoking, etc.). When an atheroma plaque ruptures, a clot typically forms and may cause thrombosis by obstructing blood flow through an artery. The clot may also migrate or form upstream (in the heart chambers for example).

High blood pressure is a major risk factor for various cardiovascular pathologies including atherosclerosis, stroke and aneurysms.

Vascular pathologies are generally associated with remodeling of the affected arterial wall. The structural and functional modifications of the wall therefore result in changes in mechanical properties. The evaluation and understanding of these changes in mechanical behavior are used to guide therapeutic solutions and prevention of cardiovascular events.

In the following section, we briefly present the evolution of the arterial structure due to aging as well as two important pathologies: aneurysm of the abdominal aorta (AAA) and hypertension.

**10.2.9.1. Vascular aging**

Aging is accompanied by significant changes in the cellular and extracellular components of the arterial wall. It has often been confused with atherosclerosis because atherosclerosis itself is strongly influenced by age. However, in contrast to atherosclerosis, which is a localized disease and results in a reduction in arterial lumen (stenosis), aging is a diffuse physiological process that leads to enlargement of arterial lumen [27]. This enlargement often occurs with an increase in blood pressure (without known causal relationship) and thus, an increase in circumferential wall stress. The internal and external arterial diameters increase during aging and the wall thickens with an important hypertrophy of the intima. The length of the arteries also increases with age and arterial tortuosities may appear [28].

From a histological point of view, there is an evolution of the composition of the ECM during development, maturation and then aging. The elastic fibers are gradually altered, they disorganize and appear finer and fragmented. The amount of collagen in the media increases with age while the elastin content remains stable. Thus the collagen to elastin ratio increases, which leads to stiffer arteries [29].

### 10.2.9.2. Hypertension

Hypertension or high blood pressure (HBP) is the most common cardiovascular disease, affecting a quarter of the global population. It is defined by a SBP greater than 140 mmHg and a DBP greater than 90 mmHg. The short-term response of a large artery to an increase in pressure consists in a distension of the arterial wall and a decrease in its thickness. The circumferential wall stress is then increased. To normalize the circumferential stress, adaptive phenomena take place in the medium term. An increase in internal diameter and thickening of the media are observed mainly by hypertrophy and hyperplasia of SMCs, and by an increase in the synthesis of collagen, elastin and proteoglycans [30, 31]. This type of remodeling has been observed both in experimental studies in animal models and in clinical studies [32]. Interestingly, despite the increase in elastin and collagen content, their relative mass densities and elastin-to-collagen volume ratios are not altered [33]. Peripheral resistance arteries may exhibit variable levels of hypertrophy and remodeling [34]. Remodeling contributes to the increase in peripheral vascular resistance generally associated with hypertension. This increase in resistance leads to an increase in pulse pressure, which in turn is an arterial remodeling trigger leading to enlargement and thickening of the large arteries [3]. Therefore, the remodeling induced remodeling by initial HBP is amplified and maintains HBP, thus forming a vicious circle [35].

Hypertension is a factor in the development of atherosclerosis [36] and aortic aneurysms [37]. This is related to multiple aspects, including increased wall stress and increased transmural diffusion. HBP is the main preventable risk factor for AAA and is treatable by appropriate management.

### 10.2.9.3. Abdominal aortic aneurysms

An aneurysm is defined as a localized and permanent dilatation of more than 50% of the initial diameter of an artery, with loss of parallelism of the edges. The aorta is a common site for the development of aneurysms, including the ascending, descending and abdominal regions. Ninety percent of the abdominal aortic aneurysms (AAA) are located between the renal arteries and the iliac bifurcation. The dilatation of the aorta gradually weakens the wall, which can then rupture and cause massive internal bleeding in more than 65% of cases. Other complications include arterial thrombosis and embolism. AAA is the third leading cause of cardiovascular mortality in developed countries and ranks as the $13^{th}$ most common cause of death in the United States [38]. Due to an aging population, increased prevalence of hypertension, increased smoking, screening programs and improved diagnostic tools, the incidence of AAA has continued to grow during the last decades. AAA more frequently affects men over 60 years of age. The prevalence after age 60 is 4 to 8% in men and 1-3% in women [37-40].

The majority of AAA are asymptomatic and go unnoticed because of their slow development. Approximately one-third of AAAs are detected in routine medical examinations by detection of pulsatile abdominal wall. Most of the time, the other two-thirds are accidentally discovered during medical imaging tests performed for other pathologies. Conventional diagnostic and surveillance techniques include abdominal ultrasound, a technique of choice because it is very sensitive, CT and MRI [41]. The growth rate of AAA is highly variable and requires regular monitoring of diameter when detected. Indeed, some aneurysms evolve very slowly and never reach the rupture point, while others may develop rapidly and reach a high risk status.

The exact causes of AAA are not yet well known. Nevertheless, several risk factors are involved in their occurrence:

- HBP [37],

- Smoking or smoking history [42],

- Family history suggesting genetic factors [44],

- Age, sex [37] and ethnicity (much lower prevalence in Asian populations) [43],

- Atherosclerosis [45],

- Obesity [46].

The development of AAA is determined by the destruction of the aortic media due to the alteration of the proteins of the ECM in the aortic wall and the loss of SMCs. One of the most important histological features of the aneurysm [47] is the decrease in the concentration of elastic fibers during the growth of AAA up to rupture. Compared to healthy abdominal aorta, a loss of up to 90% of elastin can be observed in AAA, with the remaining amount being mostly fragmented. The rarefaction of elastin seems to be one of the first steps in the formation of AAA [48]. Collagen is also involved in the pathophysiology of AAA. Unlike elastin, the collagen concentration does not decrease during the progression of AAA but tends to increase to compensate for the loss of elastin. The synthesis of collagen is increased in the adventitia which thickens by formation of perivascular fibrosis. This strengthening provides mechanical strength that compensates for the destruction of the media and its loss of functionality. Nevertheless, despite compensatory synthesis, the collagen appears altered with loss of functionality, realignment towards the circumferential direction and an increase in the degree of crosslinking which alters the aortic distensibility. Finally, degradation of collagen appears to be the ultimate cause of rupture [48].

Inflammation is a major characteristic of AAA, affecting mainly the external part of the media and the adventitia. The AAA wall has a characteristic inflammatory infiltrate composed of macrophages, T and B lymphocytes, neutrophils and mast cells. These cells lead to the release of pro-inflammatory cytokines and proteases, responsible for the degradation of ECM proteins.

To the scarcity of the ECM in the media, we must add the reduction of the density of the SMCs. Indeed, the number of SMCs in the AAA wall is reduced by 74% compared to the normal aorta. Moreover, the remaining SMCs predominantly have an advanced apoptotic state [49]. However, medial SMCs are an important source of elastin and collagen synthesis and play a protective role against inflammation and proteolysis [50]. Loss of SMCs could therefore be a major factor in the progression of AAA by unbalancing the remodeling of the ECM.

Finally, in approximately 75% of the patients, the wall of the AAA is covered with an intra-luminal thrombus (ILT) consisting of coagulation of a dense network of fibrin trapping red blood cells, platelets, blood proteins and cellular debris. It generally has an inhomogeneous structure with a thin, rigid, highly biologically active layer, while the medial and ab-luminal layers are less rigid and resistant because they are affected by progressive fibrinolysis [51]. Many studies have suggested that thrombus may play a protective role for the wall of the aneurysm by reducing and modifying the distribution of wall stresses [52-54]. However, several studies have also clearly demonstrated that the thrombus does not reduce the pressure on the aneurismal wall [176, 177]. In addition, the thrombus presents an important source of proteolytic and inflammatory factors. Comparative histological studies have shown that the thrombus-coated aneurysm wall is thinner, more inflammatory and contains less elastin and SMCs [55]. The size of the thrombus is also associated with faster AAA growth. Thus, embrittlement of the wall under the effect of the thrombus might be predominant compared to the protection by reduction of the wall stresses [56] or strains [177].

Current widespread clinical thinking is that AAA rupture is best predicted by monitoring its maximum diameter; specifically, that the risk of rupture is highest when the aneurysm reaches 5 or 5.5 cm in diameter. Such clinical assessment methods to evaluate AAA rupture potential are unreliable [188-190]. In general, an enlarging AAA is accompanied by both an increase in wall stress and a decrease in wall strength, and both of these parameters are critical and need to be taken into account as the instant of AAA rupture occurs when the former exceeds the latter. For these reasons, much attention has been focused over the years on the biomechanics of AAA, particularly with regards to wall stress assessment [190]. Laplace's law has been erroneously applied and is not reliable for the analyses of the complexly-shaped AAA [188]. Rather, more established and accurate methods such as finite element analysis (FEA) is required. Constitutive models for AAA wall and ILT continue to be developed [191]. These efforts, along with the advent of more accurate imaging techniques will lead to improved estimates of AAA wall stress and strength distributions in-vivo.

**10.2. Experimental techniques to study the mechanical behaviour of arteries**

Biomechanical properties of excised aortas, especially from animals, have been intensively studied by many researchers over the past decades [64]. Arteries exhibit a highly nonlinear mechanical behavior, which has been investigated from different point of views. As a first approach to arterial biomechanics, the macroscopic response of the arterial wall has been characterized through different uniaxial, biaxial and tension-inflation tests along different physiologically-relevant directions. More recently, experimental setups coupling mechanical testing with live microscopy have been developed to decipher the microstructural mechanisms that are behind the non-linear character of the response.

Experimental studies are necessary to study the alterations in mechanical properties caused by structural and functional changes in pathological arteries. In clinical follow-up, several non-invasive in-vivo tests can be performed to control blood pressure, wall remodeling and compliance. Where excision of samples is possible, further studies may be carried out with in-vitro mechanical characterization tests. However, due to the scarcity and difficulty of recovering intact and fresh human tissues, numerous in-vitro mechanical tests have been carried out on tissues derived from animal models that attempt to reproduce human pathologies. The results of these experimental tests have permitted the development of theoretical or numerical models. They have also made it possible to identify the material parameters for different types of arteries.

This section presents the main tests commonly performed in-vivo during patient follow-up and in-vitro on arterial samples.

### 10.3.1. In-vivo non invasive methods

Aging and changes in arterial mechanical properties can be controlled non-invasively by measuring arterial stiffness, arterial pressure, intima-media thickness and arterial diameter. Arterial stiffness has received particular attention in recent years as it can be considered as a tissue biomarker that represents the alterations in the arterial wall and measures the cumulative influence of cardiovascular risk factors over time [4]. Many methods of measuring arterial stiffness have therefore been developed and, for some of them, epidemiological studies have demonstrated the predictive value of arterial rigidity for cardiovascular events [97]. They are mainly based on techniques of applanation tonometry, ultrasonography and magnetic resonance imaging (MRI).

The measurement of carotid-femoral pulse wave velocity (PWV) is considered as the clinical reference method for determining aortic stiffness, because it is the simplest, most direct and most validated method in epidemiology, physiology and pharmacology. Indeed, the aorta represents the greater part of the path between the carotid and the femoral arteries. It is also the first artery at the exit of the heart and it contributes majorly to the diastolic relay of the cardiac contraction. As a mechanical parameter, PWV does not, of course, take into account the complexity of the mechanical properties of arteries, but it is simple to measure and has shown its effectiveness in the clinical follow-up of patients. Indeed, 19 studies have shown the predictive value of carotid-femoral PWV for cardiovascular events in various populations

[97]. Other anatomical sites are also of interest for the measurement of arterial stiffness, in particular the carotid artery which is frequently a place of atheromatous plaque formation.

### 10.3.2. In-vitro tissue tests

The different techniques that were developed for characterizing the incremental elastic modulus in-vivo reveal only a small part of the whole mechanical behaviour of the arterial wall. The in-vivo incremental elastic modulus is usually sufficient to understand the effects of the arterial compliance on the blood circulation (Windkessel effect), the pulse wave velocity effect and its changes with age [61, 62] and hypertension [63] for instance.

However, it is insufficient for modelling the response of the aortic wall to abnormal forces such as those applied by a stent or an endograft [93-95], or for determining the strength of the tissue and the risks of aneurysm rupture or dissections [96, 111].

More comprehensive characterizations of the arterial wall are necessary, from the stress-free state up to failure, under different combinations of uniaxial and biaxial loadings. These characterizations can only be carried out in-vitro on excised specimens.

In-vitro tests are performed on arterial samples taken away from their natural environment. Although some mechanical tests may approximate such conditions, they do not generally reproduce the complex in-vivo loading conditions, including the influence of perivascular tissues. However, they provide access to information that cannot be obtained in-vivo, such as residual stresses or the individual mechanical behavior of the different layers of the wall (intima, media, and adventitia). It is important to also note that no testing standards exist for soft tissues.

#### 10.3.2.1. Uniaxial tensile tests

Given the difficulty in collecting intact pathological arteries, the simplest mechanical test to be performed at the tissue scale, and one of the most commonly used, is the uniaxial tensile test. It can be carried out in the circumferential and longitudinal directions in order to assess the anisotropy of the tissues. The uniaxial tensile test measures the force exerted on the test piece in response to the elongation imposed on it, or vice-versa. A dog bone sample may be cut in the excised arterial segment, with the circumferential or longitudinal direction as main axis. The obtained stress-strain curve may be used to assess the stiffness and other material properties of the tissue as well as the stress and the strain at rupture.

### 10.3.2.2. Ring tests

Another simple mechanical test that can be performed on arterial tissues is the ring tensile test, which is particularly suited to the naturally cylindrical shape of the arteries. Two hooks are inserted into an arterial ring and then stretched and the force exerted on the ring is measured according to its elongation. These tests, used in pharmacology, also make it possible to determine the active mechanical properties related to the contraction of SMCs in response to vasoconstricting agents [98], in the presence or absence of an "intact" endothelium.

### 10.3.2.3. Biaxial tensile tests

To characterize the anisotropic nature of the arterial tissue, it is possible to carry out biaxial traction tests. These tissue characterization tests have been applied to the study of numerous vascular pathologies [1], including HBP and AAA. Concerning the latter pathology, in general, it appears that the aneurysmal tissue is stiffer than the healthy aortic tissue but is more prone to rupture. Moreover, several studies have shown an important heterogeneity of the mechanical properties depending on the location of the sample. As for the anisotropy of aneurysmal tissue, all studies do not agree: some have found an increased anisotropic character [99-102], while others have shown a tendency towards isotropy [103, 104].

### 10.3.3. In-vitro arterial tests (tension inflation)

The mechanical properties determined at the scale of the tissue are not sufficient to characterize the overall behavior of arteries, which are complex structures because of their geometry and their material heterogeneity. To analyze the mechanical behavior at the scale of the vascular structure, one can carry out tests of extension-inflation. These tests are more similar to the loading conditions encountered in-vivo and preserve the integrity of the vascular wall and its tubular structure. They typically consist of longitudinally stretching the artery until it reaches an extension close to the in-vivo conditions and then pressurizing it to physiological pressure levels. Alternatively, simpler closed-end, free-extension conditions may be applied [91]. It is then possible to obtain the pressure-volume or pressure-diameter relationships, as well as the corresponding longitudinal force values. These relationships can then be used to develop models of mechanical behavior, especially in the hyperelastic case [105-108]. The tensile-inflation tests evidence a salient feature of the in-vivo prestretch: for axial prestretches smaller (respectively, larger) than the in-vivo prestretch, the axial reaction force decreases (respectively, increases) when the pressure increases. However, when the axial prestretch is equal to the in-vivo pre stretch, the axial reaction force does not depend on the applied inner pressure [107, 109] and remains constant during pressurization. Similar results exist for isotonic tests, in which the axial force is kept constant but the axial prestretch varies with the applied pressure [106].

### 10.3.4. Use of full-field measurements

A common limitation to most of the techniques cited previously is the assumption of uniform material properties. This is also a limitation of most of finite element analyses in vascular biomechanics. This situation does not result from computational or theoretical limitations, but rather from the lack of experimental quantification of actual regional variations in material properties. The existence and potential importance of non-uniform properties is supported by advances in vascular mechanobiology [133], which imply that cells should be expected to build in regional variations in material properties.

To begin to address the need for new methods of quantification of regional material properties in blood vessels having complex geometry, techniques based on digital image correlation (DIC) were introduced [66, 68, 78, 84, 144, 145]. They allow for the geometric reconstruction and tracking of surface displacements, and thus calculation of strains, across local areas or even the entire artery using the concept of panoramic DIC [144].

Recent progress in inverse methods has permitted to reconstruct regional distributions of material properties from these data in different situations [146, 147]. Extensions using in-vivo imaging with magnetic resonance imaging (MRI) are also under development [148].

### 10.3.5. Effect of environmental and preservation conditions

### 10.3.5.1. Tissue collection

In-vivo arteries are subjected to residual stresses and prestretch, the amount of which depends on the organ and species under consideration; this in-vivo stress-strain state originates from the growth and remodeling processes undergone by arteries and allows for achieving a homeostatic stress state that is nearly uniform and equibiaxial across the arterial wall thickness [122]. As a consequence, excising and cutting open arterial segments partially release this existing in-vivo stress-strain state, but there is no certainty that the load free configuration corresponds to a stress - or strain - free configuration.

The biological nature and the physiological functions of the arterial tissue render the characterization of the mechanical properties a delicate task: harvesting the tissue implies the death of cells, with different consequences: 1) in-vitro characterization only investigates the passive response of the tissue, and therefore cannot account for the active role of smooth muscle cells in distributing the load across the tissue thickness; and 2) the end of constituent turnover and the progressive degradation of the organic constituents making up the tissue have consequences on the mechanical response of the tissue. In addition, harvesting the tissue also implies the relaxation of some of the prestress and prestretch existing in the arterial tissue in-vivo, leading to fiber rearrangements in the microstructure with consequences on the macroscopic mechanical response.

**10.3.5.2. Tissue storage**

Due to the biological nature of arterial tissue, the storage conditions and the temperature of test may have an impact on the in-vitro mechanical response of arterial tissue.

Regarding storage procedures, the arteries are usually harvested during surgery for humans, or in freshly sacrificed animals, or shortly after death. After excision, the arterial sample may be stored for 2-3 days in a saline solution at 4°C, or kept frozen at -20°C or at -80°C [123, 124], to prevent the sample from drying up, and avoid accelerated sample degradation. The impact of these different protocols on the mechanical properties was investigated by comparing the uniaxial tensile response of specimen stemming from the same tissue sample, but subjected to different preparation protocols. Regarding the temperature of test, two choices are classically made: either ambient room temperature or physiological temperature. Again, the uniaxial tensile responses relative to different temperatures of tests were compared.

Several studies [125-127] reported no significant variations in the mechanical response after storage, while, in other studies [128-130], the variations in the mechanical behavior encompassed variations in the initial and final stress-strain slopes, as well as changes in the knee point of stress strain curves, and in the ultimate stress. These variations may be explained by some damage experienced by the samples during freezing or refrigerating: formation of ice crystals, bulk water movement [128] can induce fiber cracking, loss of crosslinks, networks disruption, and cell death. These variations may also be explained by the decrease in the collagen content after 48h cold storage [129], as well as by the exact procedure followed to freeze the sample. Still, it seems impossible to decide on the directions of variations, since different studies have come to apparently contradictory results. It is however generally admitted that the mechanical properties of arterial samples are better preserved by freezing than refrigeration [129, 130]. Zemánek et al. [125] studied the influence of the temperature on the mechanical response of the arterial wall and showed that samples were stiffer at ambient temperature than at in-vivo temperature, since a temperature increase by 1°C resulted in a 5% stiffness decrease; this finding is in good agreement with [16].

### 10.3.5.3. Tissue preconditioning

Another important feature of arterial mechanics is the existence of a transient mechanical response: during the first mechanical cycles, the mechanical response of the arterial wall exhibits an important hysteresis whereby the loading and unloading paths do not coincide. The hysteresis is reduced after several load cycles; the stabilized mechanical response barely shows any hysteresis. Experimentalists usually get rid of this transient response by performing several preconditioning cycles. The number and amplitude of preconditioning cycles vary in the literature, owing to the absence of standards or guidelines, but the loading path and maximum load generally coincide with the last applied loading. The underlying microstructural mechanisms occurring during preconditioning have not been yet elucidated. They may be related to some viscous effects, since the transient response is also observed after a prolonged stop of the mechanical loading. Interestingly, Zemánek et al. [125] noticed that no preconditioning was necessary for equibiaxial tensile tests of arterial tissue.

### 10.3.5.4. Effect of strain rate

Another much debated feature is the strain-rate dependence of the arterial mechanical response. It is now admitted that at low strain rates, the mechanical response of arteries does not vary with the strain rate [112, 116, 125]. The viscous character of arteries is however a much more complex question, since creep and relaxation phenomena should be investigated.

### 10.3.6. Layer specificity

Arteries exhibit a layer-specific mechanical response, which depends on the layer morphology. Mechanical tests on the tunica intima were performed by [21, 119]: the intima exhibits a stiffer mechanical response when loaded in the longitudinal direction than in the circumferential one, which is in good agreement with the longitudinal orientation of the fibers of the internal elastic lamina [121]. However, it is generally agreed, that the tunica intima barely contributes to the mechanical response of arteries. Regarding the tunica media, the circumferential direction shows a stiffer uniaxial response than the longitudinal direction, which correlates with the preferred circumferential orientation of the fiber networks in the media. By contrast, in the tunica adventitia, the uniaxial mechanical response is stiffer in the longitudinal direction than in the circumferential one [21, 119]. Finally, the larger elastin content in the media as compared to the adventitia makes the media more compliant, while the adventitia can bear larger loads.

### 10.4. Constitutive models of arteries

As already introduced in Section 1.1.2, the elastic modulus is the most commonly used mechanical parameter to describe the behavior of elastic materials. It characterizes the linear relationship between the stress (force per unit area) and strain (ratio between the total deformation to the initial dimension) during a tensile test. If a body is isotropic, its elastic modulus is the same whatever the direction of tension and it is called Young's modulus.

However, the unique microscopic composition of the arterial wall requires anisotropic and nonlinear stress-strain relationships [1,57,58] At low strains, few collagen fibers are straightened and, most of them are crimped or curled. The elastic response of elastin dominates the mechanical behaviour and the wall is relatively extensible. At high strains, most of the collagen fibers are straightened and recruited to bear the stress. The mechanical properties of collagen dominate and the wall is relatively inextensible. Authors have been able to observe the gradual recruitment of collagen fibers [92] and relate it to the stiffening behaviour of arteries [57, 58].

This very specific structure provides optimal behaviour for expansion and contraction of the vessel wall during the cardiac cycle and limits distension of the wall when exposed to extreme pressures.

As aortic tissue is a nonlinear and anisotropic material, it is not possible to obtain a unique value of elastic modulus as the latter is continuously varying with the different forces applied. The incremental modulus, or tangent modulus, which is defined as the differentiation of the stress-strain relationship in the circumferential direction, has been proposed to take into consideration the variation of the elastic modulus.

It is usually admitted that within the range of strain variations between diastole and systole, the incremental modulus in the circumferential direction, further denoted $E$, remains constant [159, 178-180]. The range of strain variations between diastole and systole is sufficiently narrow with regard to the whole range of strain variations, which backs up the linear assumption within this physiological range.

The incremental modulus may be estimated in-vivo using simple equations. A classical approach is based on Laplace's law [56]:

$$E = \Delta p \frac{R/h}{\Delta R/R} = \frac{R/h}{DC} \qquad (10.2)$$

where $R$ is the inner radius of the artery, $h$ is the thickness, $\Delta p$ is the pressure variation between diastole and systole, $\Delta R$ is the radius variation between diastole and systole and DC is called the distensibility. Mean $\Delta R$ greater than 10% have been observed in the thoracic aorta [57, 58] and the incremental modulus is about 500 kPa.

The age effect was also investigated [61,62] with incremental moduli of 330 kPa on average below 35 years, and 670 kPa on average above 35 years. Isnard et al. [63] compared the incremental moduli of the aortic arch between hypertensive and normal subjects, reporting values of 1071 ± 131 kPa versus 526 ± 0.045 kPa. To our knowledge, the local effects of perivascular tissue and of branches on the elastic properties of the aorta have never been investigated.

Several studies [112, 113] have laid emphasis on the dependence of the mechanical response on the sample location along the aortic tree. The aortic stiffness was found to be higher in distal regions as compared to proximal regions [67, 114-116], which correlated with a higher collagen content in the distal aortic regions. This property may be directly related to the difference in mechanical function between proximal and distal regions: proximal regions of the aortic tree directly receive blood from the heart and therefore need to exhibit larger damping properties, which is microstructurally conveyed by larger elastin content and more undulated collagen bundles in the proximal aorta [114].

### 10.4.2. Non linearity

The arterial wall exhibits a highly nonlinear mechanical response, which was already described in the 1880s [110]: while at low applied stresses arteries are very easily deformed, the arterial response becomes much stiffer at higher applied stresses. This nonlinear response is mostly attributed to the interactions between collagen and elastin in the tissue. By imparting compressive stresses to the collagen [117], the presence of elastin increases the collagen folding, resulting in a more compliant response of the tissue [118], since straightened collagen fibers are much stiffer than elastin fibers. As a consequence, degrading elastin leads to vessel enlargement. This results in a mechanical response being softer (i.e. more stretchable) in the lower stress regime, and stiffer in the higher stress regime [118, 119]. Elastin degradation also leads to an earlier recruitment of collagen fibers, which is conducive to the mechanical response being stiffer sooner [114, 120]. Collagen degradation leads to the disappearance of the progressive stiffening of the mechanical response, while the initial slope of the stress-strain curves remains unchanged [119].

### 10.4.3. Hyperelasticity

Most continuum models of arteries taking into account the nonlinear behavior are hyperelastic models. In a nutshell, hyperelasticity expresses the relationship between the strain energy stored in a solid and the deformations undergone by the material. It assumes that the stress depends only on the current strain state and not on the path between initial and final strain states. Although soft tissues typically do not satisfy such assumption, Fung introduced the concept of pseudo-hyperelasticity which has rendered hyperelastic models very popular in biomechanics [16]. Functional forms of strain energy density are numerous [65]. Let us introduce the basic mathematical concepts needed to define strain energy density functions.

**10.4.3.1. Theory of finite deformations**

Deformations are mathematically described as functions which map the material particle position vector $\boldsymbol{X}$ (in the reference configuration at a given time $t$) to the material particle position vector $\boldsymbol{x}$ in the current configuration:

$$\boldsymbol{x} = \boldsymbol{\phi}(\boldsymbol{X}, t) \quad (10.3)$$

$$\boldsymbol{X} = \boldsymbol{\phi}^{-1}(\boldsymbol{x}, t) \quad (10.4)$$

The displacement vector is defined as: $\boldsymbol{U}(\boldsymbol{X}, t) = \boldsymbol{x} - \boldsymbol{X} = \boldsymbol{\phi}(\boldsymbol{X}, t) - \boldsymbol{X}$.

The velocity vector is defined by taking time derivatives of the mapping as follows,

$$\dot{\boldsymbol{U}}(\boldsymbol{X}, t) = \left[\frac{\partial \boldsymbol{\phi}(\boldsymbol{X}, t)}{\partial t}\right]_{\boldsymbol{X}} \quad (10.5)$$

It can also be expressed in terms of the spatial description by inserting the inverse mapping,

$$\boldsymbol{v}(\boldsymbol{x}, t) = \dot{\boldsymbol{U}}(\boldsymbol{\phi}^{-1}(\boldsymbol{x}, t), t) \quad (10.6)$$

The deformation gradient for the transformation is,

$$\boldsymbol{F}(\boldsymbol{X}, t) = \frac{\partial \boldsymbol{\phi}(\boldsymbol{X}, t)}{\partial x} \quad (10.7)$$

The right Cauchy Green and left Cauchy Green stretch tensor are respectively:

$$\boldsymbol{C} = \boldsymbol{F}^T \boldsymbol{F} \quad (10.8)$$

$$\boldsymbol{B} = \boldsymbol{F} \boldsymbol{F}^T \quad (10.9)$$

The Green-Lagrange strain tensor is:

$$E = \tfrac{1}{2}(C - 1) \qquad (10.10)$$

where **1** is the identity tensor.

### 10.4.3.2 Stress tensors

The first stress tensor that has to be introduced is the Cauchy stress tensor **σ**. It describes the stress state in the deformed body and is defined in the spatial configuration. The traction vector obtained from the application of the surface normal **n** is called the Cauchy traction vector:

$$t = \sigma \cdot n \qquad (10.11)$$

Since it describes the actual stress in the body, the Cauchy stress is called the true stress in engineering.

It is convenient to define the second Piola-Kirchhoff stress tensor **S** such as:

$$S = JF^{-1} \cdot \sigma \cdot F^{-T} \qquad (10.12)$$

where $J = \det(F)$. The second Piola-Kirchhoff stress tensor is defined in the reference configuration and is the work conjugate of the Green-Lagrange strain tensor **E**. It also has the attractive property that, like the Cauchy stress tensor, it is symmetric for non-polar materials.

### 10.4.3.3. Constitutive equations

In hyperelasticity, the existence of a strain energy density function $W$ is assumed from which a constitutive relation between stress and strain is derived. The total energy that is needed to deform the body is only dependent on the initial and the end states, and not on the loading path. The strain energy density function can be expressed as a function of the Cauchy Green right tensor: $W(C)$.

The stress strain relationship is written as:

$$\sigma = 2J^{-1} F \cdot \frac{\partial W}{\partial C} \cdot F^T \qquad (10.13)$$

For isotropic models, $W$ depends only on the 3 first invariants of **C**:

$$I_1 = \mathrm{tr}(C) \qquad (10.14)$$

$$I_2 = \frac{1}{2}[(\operatorname{tr}(\boldsymbol{C}))^2 - \operatorname{tr}(\boldsymbol{C}^2)] \quad (10.15)$$

$$I_3 = \det(\boldsymbol{C}) \quad (10.16)$$

For an incompressible material, the strain energy density function does not depend on $I_3$. In this case, the stress is derived according to:

$$\boldsymbol{\sigma} = -p\mathbf{1} + 2\boldsymbol{F}\cdot\frac{\partial W}{\partial \boldsymbol{C}}\cdot\boldsymbol{F}^T \quad (10.17)$$

where $p$ is a Lagrange multiplier enforcing the incompressibility constraint. In problems that can be solved analytically, $p$ is determined using boundary conditions. In numerical implementations, such as using the finite element method, $p$ is a numerically large factor that also enforces the incompressibility constraint, but through a penalty method.

Numerous energy functions have been used to describe the mechanical behavior of arteries. For a detailed review, the reader can refer to the book by Humphrey [1]. Despite the large number of data showing that healthy and aneurysmal arteries are anisotropic [102], many studies still use isotropic mechanical behavior relationships, particularly in numerical finite element models.

Among all the strain energy density functions proposed for arteries, we will distinguish those that have been established through a phenomenological approach based on experimental data, and those motivated by the arterial microstructure.

The most common strain energy density functions for arteries are:

#### 10.4.3.3.1. Demiray

$$W = \frac{\beta}{2\alpha}\left(e^{\alpha(I_1-3)} - 1\right) \quad (10.18)$$

where $\alpha$ and $\beta$ are material constants.

The Demiray strain energy function can sometimes be completed with a Neo-Hookean term:

$$W = \mu_1(I_1 - 3) + \frac{\beta}{2\alpha}\left(e^{\alpha(I_1-3)} - 1\right) \quad (10.19)$$

#### 10.4.3.3.2. Yeoh

$$W = \sum_n c_n(I_1 - 3)^n \qquad (10.20)$$

where $c_n$ are material constants.

**10.4.3.3.3. Fung**

$$W = \frac{c}{2}\left(e^{c_1 E_{\theta\theta}^2 + c_2 E_{LL}^2 + 2c_3 E_{\theta\theta} E_{LL}} - 1\right) \qquad (10.21)$$

where $c$, $c_1$, $c_2$ and $c_3$ are material constants.

**10.4.3.3.4. Holzapfel**

$$W = \frac{c}{2}(I_1 - 3) + \frac{k_1}{2k_2} \sum_{i=4,6}\left(e^{k_2(I_i - 1)^2} - 1\right) \qquad (10.22)$$

where $c$, $k_1$ and $k_2$ are material constants, $I_4 = \boldsymbol{M_1} \cdot (\boldsymbol{C} \cdot \boldsymbol{M_1})$ and $I_6 = \boldsymbol{M_2} \cdot (\boldsymbol{C} \cdot \boldsymbol{M_2})$

$\boldsymbol{M_1}$ and $\boldsymbol{M_2}$ are unit vectors characterizing two preferred directions of fibers in the reference configuration.

This strain energy function models a composite material made of a matrix reinforced with fibers. The parameters $c$ and $k_1$ are the effective stiffnesses of a matrix and of fiber phases, respectively, both having dimensions of force per unit length. The $k_2$ parameter is a non-dimensional parameter that governs the tissue's strain stiffening response.

**10.4.3.3.5. Gasser with dispersion**

$$W = \frac{c}{2}(I_1 - 3) + \frac{k_1}{2k_2} \sum_{i=4,6}\left(e^{k_2[\kappa(I_1 - 3) + (1 - 3\kappa)(I_i - 1)^2]} - 1\right) \qquad (10.23)$$

where $c$, $k_1$ and $k_2$ are material constants, and $\kappa = \int_0^\pi \rho(\theta) \sin^3(\theta) d\theta$.

When $\kappa = 0$ the equation models a composite with all the fibers perfectly aligned in the direction $\boldsymbol{M_1}$ or $\boldsymbol{M_2}$, while when $\kappa = 1/3$, the fibers would have no preferential direction (isotropic).

**10.4.3.3.6. Humphrey with 4 families of fibers**

$$W = \frac{c}{2}(I_1 - 3) + \sum_{i \in [1,4]} \frac{c_i}{2k_i}\left(e^{k_i(I_i - 1)^2} - 1\right) \qquad (10.24)$$

where $c$, $c_i$ and $k_i$ are material constants, $I_i = \boldsymbol{M_i} \cdot (\boldsymbol{C} \cdot \boldsymbol{M_i})$ and $\boldsymbol{M_i}$ is a unit vector characterizing the preferred direction of fibers in the reference configuration. The four preferred directions are the circumferential direction, the axial direction, and two symmetric diagonal directions defined by the angle to the circumferential direction $\mp\theta$.

Except Fung's model, all these models are written in their incompressible versions. They may be used also in compressible versions in which it is common to additively split the strain energy function into a term depending only on the change of volume, and another term independent of volume changes [65]. In this case, we introduce

$$\overline{F} = \frac{1}{J^{1/3}} F \qquad (10.25)$$

Then then derive $\overline{C}, \overline{E}$ and the normalized invariants as well. However, there are known issues with slightly compressible anisotropic formulations using this approach [181, 182].

**10.4.4.3 Multilayer models (including residual stresses)**

Most constitutive relations and stress analyses have employed an informal homogenization procedure and treated the wall as a single layer. Such an approach has enabled significant advances, including the discovery of important implications of residual stresses and the development of growth and remodeling models that capture salient aspects of arterial adaptation. Nevertheless, stress analyses that account for the different layers of the arterial wall, notably the media and adventitia in all vessels and also the intima in aging and particular diseases, can provide additional information that is essential depending on the question of interest. Indeed, given the recent recognition of the differential mechanobiological roles of medial smooth muscle cells and adventitial fibroblasts [166], there is a pressing need to better understand the layer-specific differences in the local mechanical environments experienced by these different types of cells.

Several investigators have used a two-layer model (the two layers being the media and the adventitia layers) with Holzapfel's model [65]. Recently, Bellini et al. [167] presented a new approach for modelling layer-specific arterial wall mechanics that was motivated by recent growth and remodeling simulations and based on histologically and clinically measurable data. In particular, in contrast to classical approaches that employed either an intact or a radially cut traction-free configuration as a reference, they used the in-vivo homeostatic state as a biologically and clinically relevant reference configuration and built a new bi-layered model of the arterial wall. Moreover, they endowed the primary structurally significant constituents—elastic fibers, smooth muscle, and fibrillar collagen—with individual ''deposition stretches'', which ensured that the in-vivo reference configuration was defined by homeostatic stresses. Embracing the material nonuniformity of the wall clearly distinguished between a requisite, computationally convenient reference configuration for the artery, and the actual stress-free (i.e., natural) configurations for the individual constituents. This ''constrained mixture'' approach allowed one to naturally account for tensile stresses in all constituents at physiologic and supra-physiologic pressures, as well as for most compressive stresses that necessarily emerge in some constituents at sub-physiologic pressures. The separate prescription of material properties in the media and adventitia was based primarily on histological information on individual mass fractions and orientations of constituents. In addition to achieving fits to in-vitro biaxial mechanical data that were comparable to prior reports that use classical homogenized models, Bellini's model also allowed one to predict associated traction-free configurations, which can serve as independent validations. Therefore, one advantage of this approach is that it does not require one to prescribe residual stress related opening angles, which cannot be measured in-vivo and cannot be prescribed easily whenever the geometry is not cylindrical. Rather, one merely needs to prescribe point-wise deposition stretches within an assumed homeostatic state, regardless of the overall geometry of the wall. A 3D finite-element implementation of this model was recently published [170].

**10.4.4. Multiscale modelling**

Although most of the models discussed previously were motivated by the structure of the arterial wall, many are phenomenological in the sense that material parameters have to be calibrated using experimental tests.

True multiscale modelling would consist in entering the mechanical properties of each micro-constituent of the arterial wall in a model and deduce the macroscopic properties using homogenization approaches.

The arterial wall owes its main mechanical characteristics, such as the progressive stiffening and anisotropy, to collagen fibers and their orientations. In most of the available constitutive models, fiber families are characterized by their orientation angles while their progressive stiffening is modeled through exponential functions of the stretch. The selection of a constitutive behavior implies the choice of a number of collagen fiber families. The determination of their orientations can be done in two ways: either by histological examination of the tissue [183], or by inverse method, searching for the orientation angles that best fit the macroscopic behavior of the tissue.

Modeling the progressive recruitment of collagen fibers is another important question that needs to be addressed. In the (ex vivo) load-free configuration, microscopic observations evidence crimped fibers with different orientations that the mechanical loading tends to stretch and reorient along the principal strain directions. At high stretches, the collagen fibers are perfectly straight and parallel to each other. However, the physiological load lies between these two extreme situations and poses the question of the collagen fiber engagement under physiological conditions. Different experimental studies have shown that only partial engagement of the collagen fibers is reached at physiological pressure: only 5-10% of the fibers actively participate in the mechanical behavior of vascular tissues at these pressures. This progressive recruitment is the physical origin of the non-linear character and progressive stiffening of the response of vascular tissues; it is generally implicitly accounted for by the introduction of exponential functions in the constitutive models (see Section 3.3.3) but in some specific models, a probability distribution function for the engagement strain of the fibers has been introduced [92].

### 10.4.5. Mechanobiology

Many experiments have shown that the stress field dictates, at least in part, the way in which the microstructure of arteries is organized. This observation led to the concept of functional adaptation wherein it is thought that arteries functionally adapt so as to maintain particular mechanical metrics (e.g., stress) near target values. To accomplish this, tissues often develop regionally varying stiffness, strength and anisotropy. Models of growth and remodeling necessarily involve equations of reaction-diffusion. There has been a trend to embed the reaction-diffusion framework within tissue mechanics [3, 4]. The primary assumption is that one models volumetric growth through a growth tensor $\mathbf{F_g}$, which describes changes between two fictitious stress-free configurations: the original body is imagined to be fictitiously cut into small stress-free pieces, each of which is allowed to grow separately via $\mathbf{F_g}$, with $\det(\mathbf{F_g})=1$. Because these growths need not be compatible, internal forces are often needed to assemble the grown pieces, via $\mathbf{F_a}$, into a continuous configuration. This, in general, produces residual stresses, which are now known to exist in many soft tissues. The formulation is completed by considering elastic deformations, via $\mathbf{F_a}$, from the intact but residually stressed traction-free configuration to a current configuration that is induced by external mechanical loads.

The initial boundary value problem is solved by introducing a constitutive relation for the stress response to the deformation $\mathbf{F_e F_a}$, which is often assumed to be incompressible hyperelastic, as well as a relation for the evolution of the stress-free configuration via $\mathbf{F_g}$. Thus, growth is assumed to occur in stress-free configurations and typically not to affect material properties. Although such theory called the theory of kinematic growth yields many reasonable predictions, Humphrey and coworkers have suggested that it models consequences of growth and remodeling, rather than the processes by which they occur. Growth and remodeling necessarily occur in stressed, not fictitious stress-free, configurations, and they occur via the production, removal, and organization of different constituents; moreover, growth and remodeling need not restore stresses exactly to homeostatic values.

Therefore, Humphrey and coworkers introduced a conceptually different approach to model growth and remodeling, one that is based on tracking the turnover of individual constituents in stressed configurations (the constrained mixture model [5, 6]). Recently, Cyron et al. [171] worked on a unified theory between kinematic growth and constrained mixture theory.

### 10.4.6. Identification of constitutive properties

In order to identify the material parameters corresponding to a given type of artery, the data from the experimental tests must be compared to the chosen mechanical behavior model by solving an *inverse problem*. In the case of conventional uniaxial or biaxial tests at the tissue scale, the experimental data can be translated directly into terms of stress-strain relationships. Therefore, it is sufficient to optimize the parameters of the constitutive model by suing the experimental stress-strain curves and minimizing the difference between the experimental stress and the stress computed by the model at the different deformation levels. Numerical nonlinear fitting methods, such as the least squares method, are used for this purpose.

In the case of more complex tests, two possibilities may arise: if simplifying hypotheses can be made to establish analytical equations of the mechanical problem, then these equations may be solved numerically (by optimization as before). This is the case, for example, of the inflation-extension test [65]. The same approach can also be used in the case of pressure-diameter data obtained in-vivo [131, 132]. If the geometry or loading is too complex to express the analytical equations of the mechanical problem, then a finite element model coupled with an inverse method may be used. Interestingly, a new approach for the bi-axial characterization of in-vitro human arteries was recently proposed [66] that permits the identification the material constants in Holzapfel's or Humphrey's models even with heterogeneous strain and stress distributions in arterial segments. From the full-field experimental data obtained from inflation/extension tests, an inverse approach, called the virtual fields method (VFM), can be used for deriving the material parameters of the tested arterial segment. The results obtained in human thoracic aortas are promising [71, 78]. More details about this method can be found in [146] and [172].

**10.5. Mechanics of the ascending thoracic aorta**

The human ascending thoracic aorta and thoracic arch are segments of an elastic artery whose mechanical properties play a crucial role in damping the pressure wave that occurs within the vessel and in channeling the blood flow coming from the heart.

The main pathologies affecting the ascending thoracic aorta are dissections and aneurysms (dilatations). Aneurysms of the ascending thoracic aorta represent on tenth of aortic aneurysms, but the surgical repair of ascending thoracic aortic pathologies, especially ATAA, and type A dissections, is very complex. Conversely to AAA, of which 50 to 75 % (depending on countries) are now treated by endovascular repair (EVAR) interventions, ATAA are commonly treated by conventional surgery, with open chest, requiring heart lung machine. Note that, as treated by conventional surgery, ATAA is a large source of excised tissue, providing many biomechanical samples.

The mechanics of the ascending thoracic aorta is special, due to its crucial role on hemodynamics, but also because it has to bear the motions of the heart, which constitutes a permanent cyclic loading that is found in no other artery [174]. When compared to AAA, the literature on computational biomechanics of aortic dissections is scarce. The few reports that are available, highlight the potential of computational modeling, but also reveal many shortcomings [168]. For all these reasons, we decided to dedicate a specific section to the mechanics of the ascending thoracic aorta and ATAA.

### 10.5.1. Fracture mechanics of ATAA

Tensile strengths of the thoracic aorta in the axial and circumferential directions were previously reported in the literature [69-75,78-82]. Mohan and Melvin [77] reported an average ultimate stress of 1.14 MPa in quasi-static biaxial tension and an average ultimate stress of 1.96 MPa in dynamic ($20s^{-1}$) biaxial tension. McLean et al. [81] identified the strength in the radial direction, but on porcine data (0.06 MPa). Sommer et al. [82] showed that porcine tissues may be significantly different of human ones. Strength was also identified using inflation tests on intact segments [83] yielding an average strength of 2.5 MPa.

Compared to normal tissue, aneurysmal specimens displayed lower wall thickness and failure strain, higher maximum elastic modulus (MEM), and equal failure stress than control specimens in the majority of regions and directions [75]. ATAA formation was associated with stiffening and weakening of the aortic wall [70].

More recently, the anisotropy of the aortic wall was demonstrated by significantly different results for MEM between the circumferential and longitudinal orientations [74]. Interestingly, the aorta is stiffer longitudinally in the greater curvature than in the lower curvature, which can correlate to the physiological movement of the arch [76]. Marked heterogeneity was evident in healthy and aneurysmal aortas (variations between anterior, posterior and lateral) [72].

Uniaxial tests do not reveal the whole mechanical behavior of the arterial walls as there exists coupling effects between the axial and circumferential stresses. Biaxial tests have to be carried out for a complete characterization [77]. An interesting procedure for characterizing the biaxial properties of the aorta is the bulge inflation test combined with digital image correlation to measure the local strain and stress fields across the specimen and to derive the local constitutive parameters using an inverse approach [78]. It does not reveal any significant imbalance between the circumferential and axial directions in terms of response, which tends to show that the aneurysm does not present a marked anisotropy in terms of elastic response. This suggests that wall degeneration has led to a more random distribution of the collagen fibers.

Biomechanical studies have also achieved better insight in ATAA rupture or dissection [70, 88,156] and attempted to elucidate the risk profile of the thoracic aorta [157,12]. Recently, our group developed an approach to identify the patient specific material properties of ATAAs by minimization of the difference between model predictions and gated CT images [159]. Moreover, we characterized the mechanical properties of ATAA on samples collected from patients undergoing surgical repair. We also defined a rupture risk based on the brittleness of the tissue (the rupture criterion is reached when the stretch applied to the tissue is greater than its maximum extensibility or distensibility) and showed a strong correlation between this rupture risk criterion and the physiological elastic modulus of ATAAs estimated from a bulge inflation test [14]. A failure criterion based on in-vitro ultimate stretch showed a significant correlation with the aortic membrane stiffness deduced from in-vivo distensibility. The strength of the aortic tissue is generally defined as the maximum stress that the tissue can withstand before failing. However, when the maximum stress ratio between the stress applied to the tissue in-vivo and its strength was derived, it was noted that most of the collected ATAA samples were far from rupture. Alternatively one can define rupture when the stretch applied to the tissue exceeds its maximum extensibility or distensibility. This definition of rupture may even be more physiologically meaningful as it is reported that aneurysm ruptures or dissections often occur at a time of severe emotional stress or physical exertion [161]. Such situations can induce significant changes in blood volumes in the aorta, making less compliant aneurysms more prone to rupture as they cannot sustain such volume changes. Based on this analysis, Martin et al. defined a similar criterion, named the diameter risk, which is the ratio between the current diameter of the aneurysm and the rupture diameter [161]. They showed that the diameter risk increased significantly with the physiological elastic modulus of the artery. Indeed, if the aortic wall is stiff, a rather large increase of pressure can be induced by a small increase in blood volume.

One explanation for this phenomenon may be the mechanism of collagen recruitment described by Hill et al. [92]. As collagen recruitment increases with the load, the tangent stiffness of the tissue increases too. Collagen recruitment can be expected to be delayed in the tissue when it still contains a significant fraction of intact elastin. At physiological pressures, when elastin is not fragmented, only a small fraction of collagen contributes to the aortic stiffness, which is relatively small (below 1 MPa). Conversely, when elastic lamellae and elastic fibers are highly disrupted, collagen can be expected to be recruited earlier and to contribute significantly to the stiffness at physiological pressures, thereby increasing the stiffness (above 1 MPa).

This mechanism of higher recruitment of collagen at physiological pressures is in line with the findings of Iliopoulos et al. [75] who showed that the fraction of elastin, but not collagen, decreased in ATAA specimens, displaying lower wall thickness and failure strain, higher peak elastic modulus, and equal failure stress than control specimens. Moreover, it is commonly admitted that the elastin fraction decreases with age.

The level of collagen recruitment may explain why the stretch based rupture criterion can be very close to 1, but not the relatively high values of the stress based rupture criterion that occur for some patients. Indeed, some specimens showed a relatively large extensibility but ruptured at a relatively small stress. When collagen is recruited, it can withstand stresses until damage initiates with ruptures at the fiber level as demonstrated by Weisbecker et al. [163,164]. A dense and crosslinked network of collagen will permit to reach larger stresses before the initiation of damage and eventually ultimate stress.

**10.5.2. Fluid mechanics in ATAA**

Perturbed hemodynamics has frequently been reported in ATAA [184-186]. It is a consequence of the altered aortic arch configuration (for instance shift from type I to type II as introduced in Fig. 5) but it is often associated with aortic valve phenotype which is believed to play a key role in the development and the growth of clinically significant aortic dilatations. The bicuspid aortic valve (BAV) lesions are reported to considerably change the distribution and magnitude of the wall shear stress (WSS) along the ATAA. Moreover, recent studies have shown that even in the absence of ATAA, the aortic annulus and root dimensions are significantly larger in patients with BAV aortic insufficiency (AI) than patients with BAV stenosis [187].

Our group recently examined the hemodynamics in ATAA with concomitant aortic insufficiency (AI) by combining 4D MRI analyses and CFD studies [173]. The objectives of this research were to observe the impact of secondary flow on the ATAA hemodynamics and to understand how the degree of AI may be indicative of possible rupture risk. From the 3D streamlines, flow eccentricity was observed in patients affected by BAV and AI. During systole, flow helicity induced flow impingement against the aneurysmal wall generating a non-homeostatic distribution of the WSS which was found to be higher and more extended than (tricuspid aortic valve) TAV patients. This helicity grew during the blood flow deceleration and diastolic phase due to the blood flow regurgitation resulting in local wall shear stress (WSS) peaks. The highest time averaged wall shear stress (TAWSS) was found in these areas. From the velocity field results, recirculation and vortices were found in the ATAA region in TAV patients affected by AI. The jet flow impingement against the aortic wall was found in the region around the bulge, downstream of the area of maximum dilatation. The maximum TAWSS was found in the same region. One patient showed no region of jet flow impingement but developed a bulge on the inner curvature side of the aorta. In summary, the existence of jet flow impingement and the TAWSS magnitude at its location were mostly associated with the orientation of the inlet. This orientation was quantified as the angle between the plane of the inlet cross section and the transverse plane of the thorax. An angle of about 25 degrees exists in healthy subjects. The angle can go up to 70 degrees in patients who have the largest TAWSS. The angle change comes from the change in aorta morphology during ATAA development. ATAA patients commonly have a larger curvature which may result from a larger increase of the aorta length on the outer curvature side than on the inner curvature side. Different structural properties on the inner and outer curvature sides have already been reported and may be associated with the contact between the pulmonary artery and the aorta on the inner curvature side. It should be noted that the ATAA development is associated with a decrease of the aorta axial stretch in mice models of the Marfan syndrome, which results from elastin fragmentation and faster collagen turnover. These effects (elastin loss, axial stretch decrease) are expected to be more pronounced in the areas exposed to larger hemodynamic loading which would be more favorable to fatigue damage of elastin and aorta remodeling. Moreover, this remodeling (i.e. aorta diameter increase, arch length increase and aortic unfolding) may lead to functional aortic alterations such as decreased aortic distensibility and increased aortic arch pulse wave velocity.

It also appeared from our study that no simple relationship exists between wall strength and WSS. Large TAWSS values in the bulging region of aneurysms may even be associated with relatively large strength. In addition, large strength difference between TAV and BAV patients is associated with relatively small peak of WSS and minor TAWSS variations. Actually, if it can be showed that aortic wall remodeling and subsequent morphological changes have an important impact on WSS, the reciprocal is less obvious as is still unclear whether or not WSS-mediated mechanics affect remodeling in the ascending thoracic aorta. Generally, maintenance of arterial caliber in response to increased blood pressure (to restore WSS to normal) tends to involve vessel-level changes in vasoactivity, which is greater in muscular arteries than elastic arteries. There is yet little information on possible basal tone in the human aorta.

Some authors have shown the effect of the type of aortic valve on blood flow in the ascending thoracic aorta [165]. Patient-specific computational fluid dynamics has revealed high WSS and lower oscillatory shear index in the greater curvature of bicuspid aortic valve aortas, with highly eccentric and helical flow. In patients with aortic valve disease and aortic aneurysm, morbidity or mortality can occur before size criteria for intervention are met. Patient specific CFD provides non-invasive functional and hemodynamic assessment of the thoracic aorta. With validation, it may enable the development of an individualized approach to diagnosis and management of aortic disease beyond traditional guidelines.

### 10.5.3. Mechanics of dissections

The failure properties of greatest interest to the thoracic aorta are the dissection properties [86-88]. From the mechanical point of view, tear and dissection appear if the stresses acting on the wall rise above the ultimate value for the aorta wall tissue [88-90]. Pasta et al. [88] performed delamination tests on thoracic aorta tissue samples excised from patients undergoing ATAA repair. They manually created an initial delamination plane (i.e. an intimal tear) and showed a difference in dissection strength between the circumferential and longitudinal directions and effectively measured the separation forces. Some authors also characterized the dissection energy [83]. Reported values, for an abdominal aorta, are below 10 mJ/cm.

Yet, no direct in-vivo measurement of stresses or strength is feasible. Stresses in the aorta wall are due to the concomitant influences of many factors, including the shape of the aorta, the characteristics of the wall material and the interaction between the fluid and solid domains. Mechanical stress plays a crucial role in the function of the cardiovascular system; therefore, stress analysis is a useful tool for understanding vascular pathophysiology [91].

One of the earlier studies [86] using an idealised geometry, suggested that the longitudinal stress in the wall could be a key cause of circumferentially-orientated tears, which have been reported to be the more common orientation.

Nathan et al. [154] used FEA to demonstrate increased wall stress in the human thoracic aorta above the sino-tubular junction (STJ) and distal to the left subclavian artery (LSA), where the intimal tears that result in type A and type B aortic dissections typically occur. The wall stress above the STJ was greater than that distal to the LSA, consistent with type A dissections being more common than type B dissections. Their findings suggest that localized maxima in peak pressure load-induced wall stresses above the STJ and distal to the LSA ostium may account for the development of type A and type B aortic dissections, respectively, which commonly occur at these locations

They assessed the possible effects of aortic arch configuration on wall stress in the aortic arch. The patients were classified by location of the LSA with respect to the plane parallel to the outer (greater) curvature of the aortic arch (Fig. 5). In the first group (type I aortic arch), the LSA arose from the top, or outer curvature, of the aortic arch. In the second group (type II aortic arch), the LSA arose below the outer (greater) curvature of the aortic arch. This aortic arch classification was based on age-related changes in arch configuration: older patients more often have type II arches [155]. However, the aortic arch configuration had no effect on the distribution of predicted wall stress in the normal human thoracic aorta.

Because mechanical stress in the aortic wall is proportional to blood pressure and vessel diameter, hypertension and aortic dilation are known risk factors for dissections. Wall abnormalities may also promote dissections, but similar alterations have been reported in normal aging and may not alone cause dissections. Interestingly, these factors fail to explain the transverse orientation and the common proximal location of the tear. This is important because aortic dissection will be prevented only when its underlying causes are better identified. Beller et al. [152] proposed that aortic root motion may be an additional risk factor for aortic dissection, determining both the tear location and orientation by increasing the wall stress in a specific manner. Several studies applying cinematography and contrast injections have visualized aortic root motion wherein the root is displaced downward during systole and returns to its previous position in diastole [149]. Cine-MRI studies in healthy subjects revealed an axial downward motion of 8.9 mm [150] and a clockwise axial twist of 6 degrees during systole [151]. The force driving the aortic annulus motion is the ventricular traction accompanying every heartbeat. This force is transmitted to the aortic root, the ascending aorta, the transverse aortic arch and the supra-aortic vessels. Thus, the aortic root motion has a direct influence on the deformation of the aorta and on the mechanical stress exerted on the aortic wall.

**10.6. Conclusions**

This chapter focusing on arterial and aortic mechanics has revealed that intense research has been carried out on the measurement of mechanical properties and on how they can be used in computational models to predict the biomechanical response of arteries and the aorta in many different conditions.

However, several aspects need to be improved before computational models can reliably be used in clinical situations for diagnosis purpose or surgical planning. Such aspects include the boundary conditions and the material properties that have to be obtained patient-specifically. These parameters are difficult to obtain in real clinical situations, and have been approximated to a large extent in the literature. The rapid development of 4D MRI [173] may permit important progress in these regards in the near future.

The combination of CFD and structural FE analyses has led to a better understanding of flow pressure distribution, wall shear stress quantification, and the effect of material properties and geometrical parameters. Computational methods have made patient-specific analyses possible, a feature which is essential to deciphering the special biomechanical conditions at play in a particular ATAA patient. Each patient has their own unique anatomy and pathophysiology that affect material properties and boundary conditions, which in turn can significantly influence their clinical management.

Unfortunately, CFD/FE methods, whilst potentially providing crucial information about the pathophysiological mechanics of an aneurysm, can be very time consuming to undertake. This is made worse by the fact that fluid-structure-interactions (FSI) approaches are computationally expensive and complex methods for routine use in real-life clinical situations. Therefore, a much simpler interface needs to be developed, wherein the vascular clinician is in a position to assess the prognosis of the patient, evaluate the rupture risk of the aneurysm and proceed to surgical planning, either with endovascular or conservative treatments, but always in a customized fashion.

Finally, the most important challenge in aortic mechanics is to appropriately characterize and model the structural alterations of arteries. For instance, the growth rate of an ATAA, used as a criterion to justify surgical intervention, is calculated from maximum diameter measurements at two subsequent time points; however, this measure cannot reflect the complex changes in vessel wall morphology and local areas of weakening that underline the strong regional heterogeneity of ATAA. Indeed, ATAA disease is characterized by a strong regional heterogeneity within the thoracic segments in terms of biomechanical properties, atherosclerotic distribution, proteolytic activity, and cell signaling pathways. The current intense research efforts in mechanobiology will hopefully lead to major breakthroughs in this area in the coming years.

**Figure Legend**

Figure 10.1: Schematic representation of two successive cycles of arterial pressure.

Figure 10.2: Schematic of arterial pressure waveforms and calculation of augmentation index A) Pulse waveforms in healthy compliant vasculature, timing of rebound wave reflection occurs during diastole (D). B) Pulse wave reflection is faster and earlier. Adapted from [169].

Figure 10.3: Schematic of the opening angle experiment.

Figure 10.4: Comparison of blood flows for heathy ascending thoracic aorta and ATAA.

Figure 10.5: Aortic arch classification: (A) in the type I arch, the LSA originates from the top of the (outer) curvature. (B) In the type II arch, the LSA originates from below the outer curvature. (source: [154])

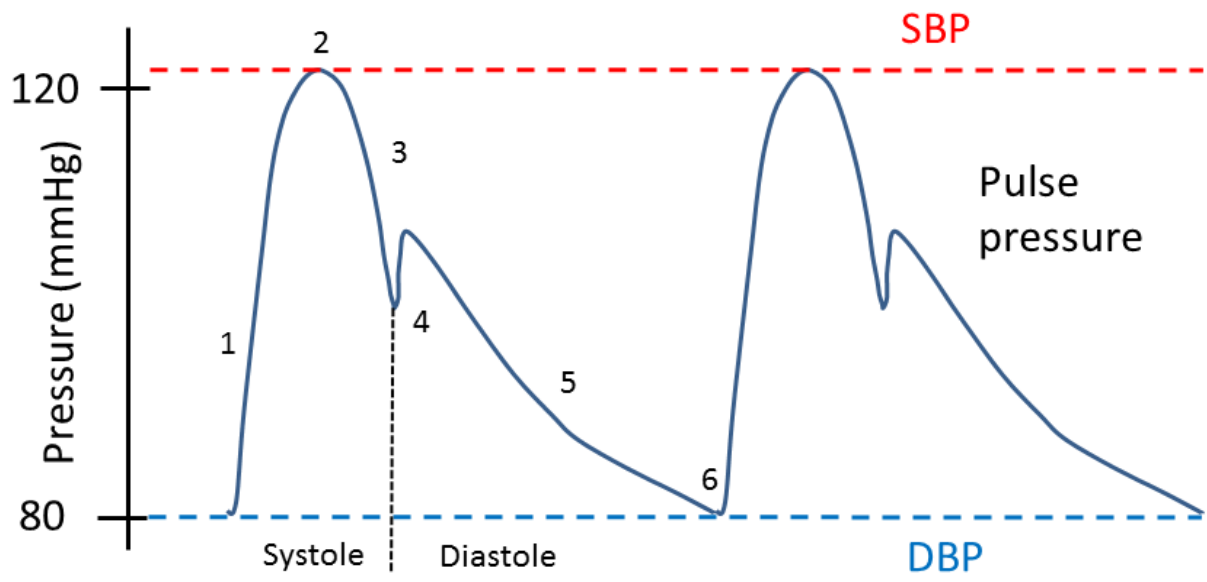

Fig. 1.

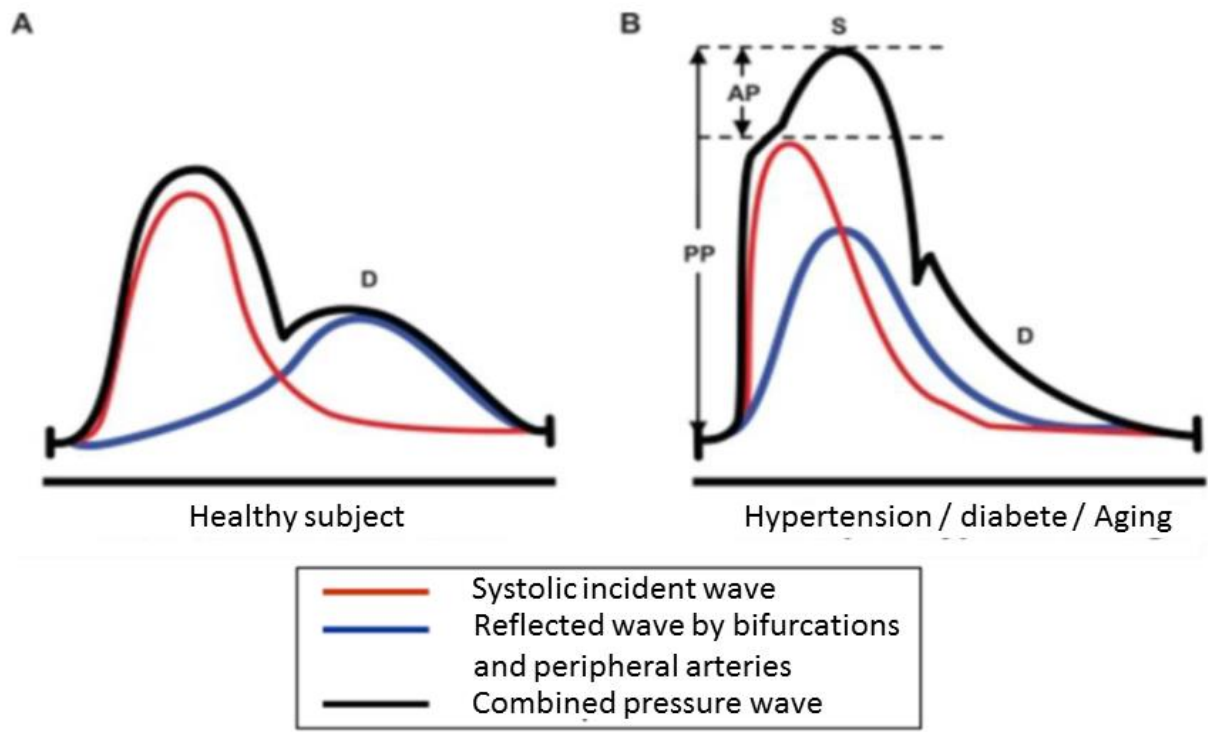

Fig. 2.

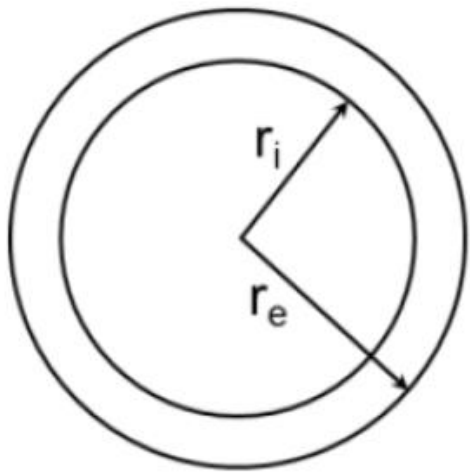 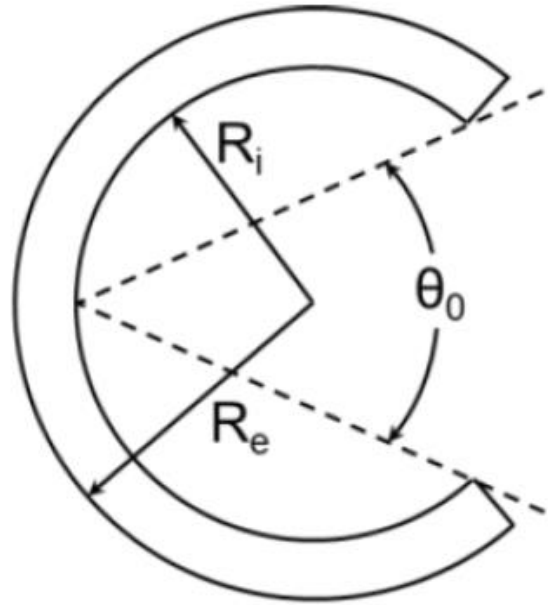

Closed    Open

Fig. 3.

| 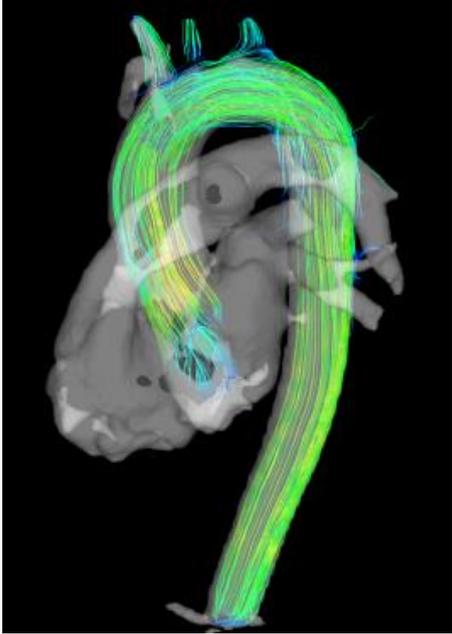 | 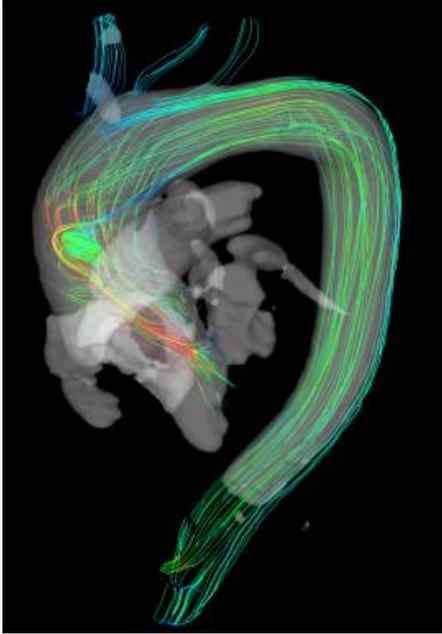 |
|---|---|
| Laminar flow in a healthy young patient | Disturbed hemodynamics in a patients suffering from bith ATAA and aortic insufficiency |

Fig. 4.

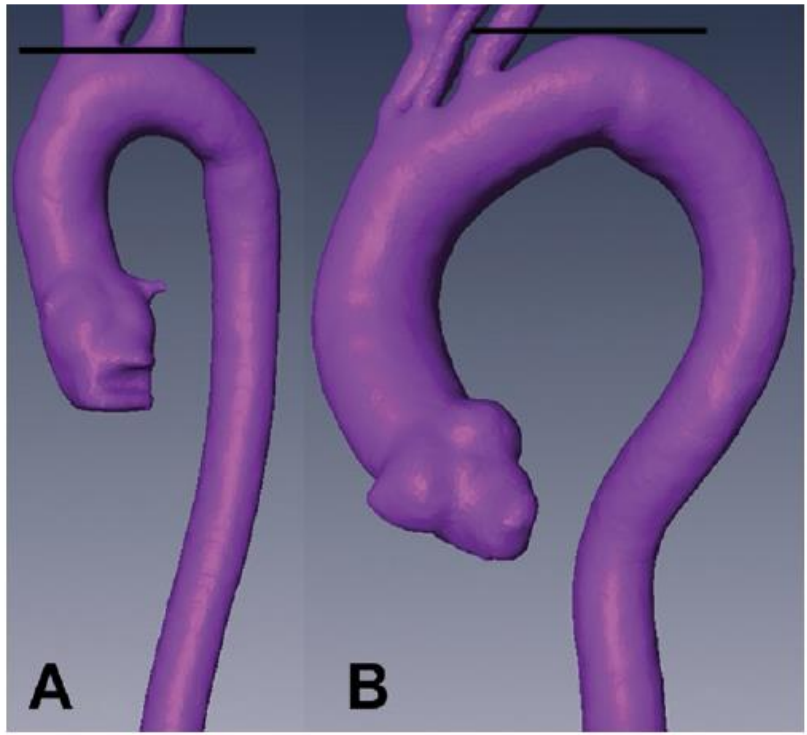

Fig. 5.